\documentclass[twocolumn,showpacs,prb,superscriptaddress]{revtex4-1}
\usepackage{amsmath}
\usepackage{amsfonts}
\usepackage{amssymb}
\usepackage{dcolumn}
\usepackage{bm}
\usepackage[dvips]{graphicx}
\usepackage{epsfig}

\begin{document}

\title{Interaction-disorder competition in a spin system evaluated through the Loschmidt Echo}

\author{Pablo R. Zangara}
\affiliation{Instituto de F\'{i}sica Enrique Gaviola (IFEG), CONICET-UNC and Facultad
de Matem\'{a}tica, Astronom\'{i}a y F\'{i}sica, Universidad Nacional
de C\'{o}rdoba, 5000 C\'{o}rdoba, Argentina}
\author{Axel D. Dente}
\affiliation{Instituto de F\'{i}sica Enrique Gaviola (IFEG), CONICET-UNC and Facultad
de Matem\'{a}tica, Astronom\'{i}a y F\'{i}sica, Universidad Nacional
de C\'{o}rdoba, 5000 C\'{o}rdoba, Argentina}
\affiliation{INVAP S.E., 8403 San Carlos de Bariloche, Argentina}
\author{An\'{\i}bal Iucci}
\affiliation{Instituto de F\'{i}sica La Plata (IFLP), CONICET-UNLP and Departamento de F\'{i}sica, Universidad Nacional de La Plata, CC 67, 1900 La Plata, Argentina}
\author{Patricia R. Levstein}
\affiliation{Instituto de F\'{i}sica Enrique Gaviola (IFEG), CONICET-UNC and Facultad
de Matem\'{a}tica, Astronom\'{i}a y F\'{i}sica, Universidad Nacional
de C\'{o}rdoba, 5000 C\'{o}rdoba, Argentina}
\author{Horacio M. Pastawski}
\affiliation{Instituto de F\'{i}sica Enrique Gaviola (IFEG), CONICET-UNC and Facultad
de Matem\'{a}tica, Astronom\'{i}a y F\'{i}sica, Universidad Nacional
de C\'{o}rdoba, 5000 C\'{o}rdoba, Argentina}
\email{horacio@famaf.unc.edu.ar}

\begin{abstract}
The interplay between interactions and disorder in closed quantum many-body
systems is relevant for thermalization phenomenon. In this article, we address
this competition in an infinite temperature spin system, by means of the
Loschmidt echo (LE), which is based on a time reversal procedure. This
quantity has been formerly employed to connect quantum and classical chaos,
and in the present many-body scenario we use it as a dynamical witness. We
assess the LE time scales as a function of disorder and interaction strengths.
The strategy enables a qualitative phase diagram that shows the regions of
ergodic and nonergodic behavior of the polarization under the echo dynamics.
\end{abstract}
\maketitle

The ergodic hypothesis of statistical mechanics implies the equivalence
between time and ensemble averages. It is expected that a conservative
many-body system satisfying such hypothesis would explore uniformly the whole
energy shell. It is now a long time since Fermi, Pasta and Ulam (FPU)
\cite{FPU} questioned how the irregular dynamics induced by nonlinearities in
a Hamiltonian may lead to energy equipartition as a manifestation of
ergodicity. Such dynamics, referred to as \textit{thermalization}, did not show
up in their pioneering numerical simulations. Even though their
striking results are now explained by the theory of
chaos\ \cite{izraChaosReview}, the solution of the quantum analogues are still
in the early stages \cite{Eisert2011,polkovnikovRMP}.

Thermalization in isolated, strongly interacting, quantum systems is defined
relative to a certain set of observables \cite{lebowitz2010}. In particular,
remarkable experiments have been performed to monitor momentum distributions
of cold atoms loaded in one-dimensional optical lattices
\cite{CradleNature2006,*BlochNatPhys2012}, where the integrability of the
underlying dynamics is weakly broken. Accordingly, a fundamental question is
whether a nonergodic to ergodic transition threshold exists as one may go parametrically from an
integrable to a nonintegrable quantum system. In the
FPU problem one can surely answer affirmatively, since the onset of dynamical
chaos can play the role for such a transition \cite{izraChaosReview}. For
interacting quantum systems in the presence of disorder, it is expected that a phase transition exists
between delocalized states and a phase characterized by many-body localization
(MBL) \cite{altshuler2006}. This would constitute the sought threshold between
ergodic and nonergodic behavior \cite{polkovnikovRMP}.

The MBL results from a quantum dynamical phase transition between diffusive
(ergodic) and localized (nonergodic) dynamics
\cite{palhuse2010,oganesyanhuse2007}. It occurs at nonzero (and eventually
infinite) temperature, and is manifested in dynamical properties. The crucial
idea is a competition between interactions and Anderson disorder
\cite{AndersonRMP-1978}. In the many-body ergodic phase, the expectation
values of observables computed on a finite subsystem, using a single energy
eigenstate of the whole interacting system, would coincide with those
evaluated in the corresponding microcanonical thermal ensemble
\cite{popescuNATURE2006,rigolNATURE2008}. In this condition one may say that
the system acts as its own heat bath. Quite on the contrary, in the
nonergodic many-body phase, dynamics resembles a \textquotedblleft
glassy\textquotedblright\ behavior that precludes self-thermalization. In
view of the difficulties of addressing a full many-body dynamics of
specific correlation functions \cite{prosen2008,pollmann2012}, much of the
progress in assessing the transition between the mentioned regimes relied
mainly on the evaluation of spectral properties
\cite{oganesyanhuse2007,Altshuler2010}.

The dynamics of specific observables has proved useful to monitor
the onset of many-body chaos \cite{Flambaum2001dynamics}. In
particular, the Survival Probability (SP) of the eigenstates of an unperturbed
Hamiltonian $\hat{H}_{0}$ under the evolution of the full interacting
Hamiltonian $\hat{H}_{0}+\hat{\Sigma}$ decays with a characteristic time scale
$\tau$. In such a case, a crossover from an exponential controlled by
$1/\tau\propto||\hat{\Sigma}||^{2}$ to a Gaussian characterized by
$1/\tau\propto||\hat{\Sigma}||$ is interpreted as evidence of the onset of a
chaotic structure in the eigenstates of $\hat{H}_{0}+\hat{\Sigma}$
\cite{Flambaum2001,marquardt2010,santos2012}. The first regime is described by
the Fermi golden rule\ (FGR), and hence the breakdown of the FGR leads to
dynamical chaos. Much in the spirit of this dynamical approach, in this
article we propose and analyze the evolution of an experimentally accessible
local observable, as a way to assess the interaction-disorder competition in a
many-spin system. We resort to the \textit{Loschmidt echo} (LE)
\cite{prosen,*Jacquod,*scholarpedia}, a measure of the revival that occurs when a time-reversal
procedure is applied to $\hat{H}_{0}$. The LE has been
experimentally\ evaluated from the local polarization of spin systems to
quantify the role of perturbations (i.e. $\hat{\Sigma}$) and the system's own
complexity on dynamical reversibility \cite{patricia98}. Besides, in
classically chaotic systems, it is well-known that the LE of a
semiclassical excitation undergoes a transition into a regime where its decay
rate is given by the classical Lyapunov exponent \cite{jalpa}. Here, we use
the LE dynamics as a time-dependent autocorrelation function that
quantifies the time scales and ergodicity. Specifically, the time reversal
procedure \textquotedblleft filters out\textquotedblright\ the
\textit{irrelevant} dynamics produced by an integrable $\hat{H}_{0}$ which
would hide the sought information \cite{nosotros2012}. Thus, the LE
becomes a privileged dynamical witness of the competition between the
interactions and disorder and a potential experimental candidate.

\textit{The spin model}.- We consider a spin chain, whose dynamics is given by
the total Hamiltonian $\hat{H}=\hat{H}_{0}+\hat{\Sigma}$:%

\begin{equation}
\hat{H}_{0}=\sum_{i=1}^{N}J\left[  S_{i}^{x}S_{i+1}^{x}+S_{i}^{y}S_{i+1}%
^{y}\right] \label{H1}%
\end{equation}

\begin{equation}
\hat{\Sigma}=\sum_{i=1}^{N}\Delta S_{i}^{z}S_{i+1}^{z}+\sum_{i=1}^{N}%
h_{i}S_{i}^{z}\label{H2}%
\end{equation}

where $\Delta$ is the magnitude of the homogeneous Ising interaction and
$h_{i}$ are randomly distributed fields in an interval $[-h,h]$. Periodic
boundary conditions (ring) are imposed and, unless explicitly stated, $N=12$.
Notice that $\hat{H}_{0}$ can be mapped into two independent noninteracting
fermion systems by the Wigner-Jordan transformation \cite{danieli-CPL2004},
while $\hat{\Sigma}$ includes both the two-body Ising interaction and the
local fields (disorder). Since $\hat{H}_{0}$ encloses single-particle physics,
we consider it as the irrelevant part of the dynamics, a term that the LE
manages to get rid of, allowing us to focus on interactions and disorder. This
fact justifies the idea of using the LE as a filter for the relevant physical processes.

\textit{The spin autocorrelation function evaluated as a (local) Loschmidt
echo}.- We consider a high (infinite) temperature state formally denoted
by $\left\vert \Psi_{eq}\right\rangle $, which represents an ensemble average
over all basis states with the same statistical weights. We study the spin
autocorrelation function at the particular site $1$,%
\begin{equation}
M_{1,1}(2t_{R})=\frac{\left\langle \Psi_{eq}^{{}}\right\vert \hat{S}_{1}%
^{z}(2t_{R})\hat{S}_{1}^{z}(0)\left\vert \Psi_{eq}^{{}}\right\rangle
}{\left\langle \Psi_{eq}^{{}}\right\vert \hat{S}_{1}^{z}(0)\hat{S}_{1}%
^{z}(0)\left\vert \Psi_{eq}^{{}}\right\rangle }\label{Autocorrelation1}%
\end{equation}
where the spin operator is written in the Heisenberg picture as:%
\begin{equation}
\hat{S}_{1}^{z}(2t_{R})=\hat{U}_{+}^{\dag}(t_{R})\hat{U}_{-}^{\dag}(t_{R}%
)\hat{S}_{1}^{z}\hat{U}_{-}^{{}}(t_{R})\hat{U}_{+}^{{}}(t_{R}).\label{spinop}%
\end{equation}

The evolution operators are $\hat{U}_{+}^{{}}(t_{R})=\exp[-\frac{\mathrm{i}%
}{\hbar}(\hat{H}_{0}^{{}}+\hat{\Sigma})t_{R}]$ and $\hat{U}_{-}^{{}}%
(t_{R})=\exp[-\frac{\mathrm{i}}{\hbar}(-\hat{H}_{0}^{{}}+\hat{\Sigma})t_{R}]$.
Therefore, it is explicit that the echo procedure performed over $\hat{H}%
_{0}^{{}}$ yields a global evolution operator $\hat{U}(2t_{R})=\hat{U}_{-}%
^{{}}(t_{R})\hat{U}_{+}^{{}}(t_{R})$. Using cyclic invariance of the trace,
one can replace the average over all basis states by an average over all
states that have spin $1$ up-polarized \cite{nosotros2012}. Additionally, we
replace the ensemble average by an average over a few entangled states
\cite{Alv-parallelism}, which provides a quadratic speedup of computational
efforts. It yields equivalent results provided that only local observables
(e.g. Eq. \ref{Autocorrelation1}) are evaluated. Thus, we consider:%
\begin{equation}
\left\vert \Psi_{neq}\right\rangle =\left\vert \uparrow_{1}\right\rangle
\otimes\left\{
{\displaystyle\sum\limits_{r=1}^{2^{N-1}}}
\frac{1}{\sqrt{2^{N-1}}}e^{\mathrm{i}\varphi_{r}}\text{\ }\left\vert \beta
_{r}\right\rangle \right\}  ,\text{ \ }\label{neqstate}%
\end{equation}
where $\varphi_{r}^{{}}$ is a random phase and $\left\{  \left\vert \beta
_{r}\right\rangle \right\}  $ are state vectors in the computational Ising
basis of the $N-1$ remaining spins. The $M_{1,1}$ is now written in the
Schr\"{o}dinger picture as:%

\begin{equation}
M_{1,1}(2t)=2\left\langle \Psi_{neq}\right\vert \hat{U}_{+}^{\dag}(t)\hat
{U}_{-}^{\dag}(t)\hat{S}_{1}^{z}\hat{U}_{-}^{{}}(t)\hat{U}_{+}^{{}%
}(t)\left\vert \Psi_{neq}\right\rangle \label{eco}%
\end{equation}

The explicit time dependence of $M_{1,1}(2t) $  is evaluated by means of a fourth order Trotter-Suzuki decomposition without any Hilbert space truncation, implemented on general purpose graphical processing units~\cite{Dente2013}. 

\begin{figure}
    \centering
     \includegraphics[width=0.5\textwidth]{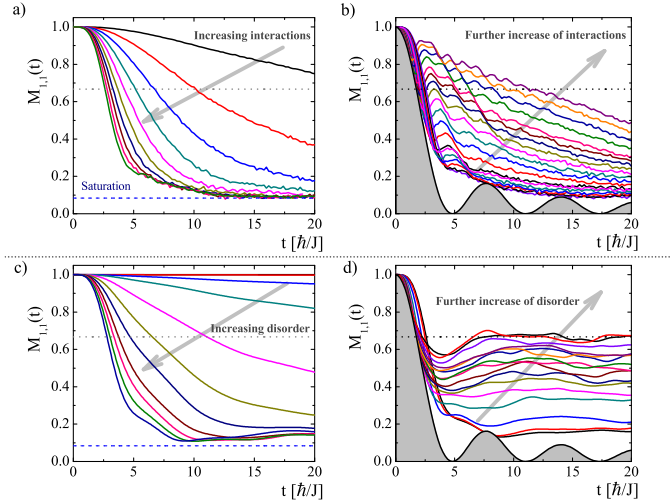}\\
      \caption{Color online. Loschmidt echo for different parameter regimes. The characteristic
time scale is defined by the decay up to $2/3$, shown by the horizontal dotted
line. (\textbf{a}) $h=0$, $0<\Delta\lesssim1.5J$. Smooth decay, until
saturation is reached (horizontal dashed line). (\textbf{b}) $h=0$,
$1.5J<\Delta<4.8J$. LE long tails destroys saturation. The SP given by Eq.
\ref{SP} determines the limit time scale (shaded region). (\textbf{c})
$\Delta=0$, $0<h\lesssim1.0J$. The localization length remains bigger than the
system size. (\textbf{d}) $\Delta=0$, $1.0J<h<5.5J$. The localization length
turns to be smaller than the system's size, and the polarization keeps around
site $1$.}%
    \label{Fig. ECOS}%
\end{figure}

\begin{figure}
    \centering
     \includegraphics[width=0.45\textwidth]{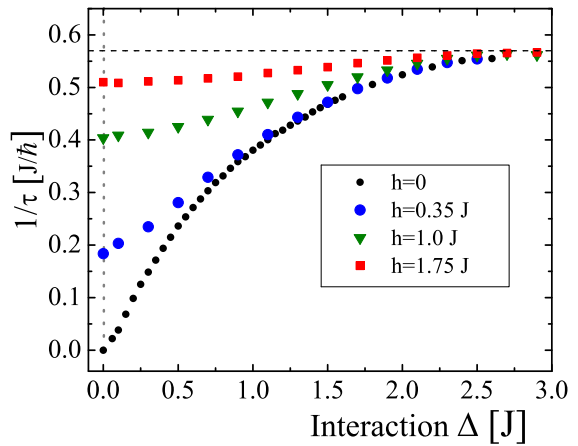}\\
      \caption{Color online. LE decay rates as a function of the interaction strength $\Delta$,
for different disorder magnitudes $h$. Horizontal dashed line represents the
($\hat{H}_{0}$) SP time scale, i.e. Eq. \ref{SP}. Data for $\Delta=0$ was
obtained from single-particle physics and 500 disorder realizations, while
data for $\Delta>0$ was obtained from full many-body simulations and 10
disorder realizations. The smoothness of matching evidences the robustness of
the simulations performed.}%
    \label{Fig. Rates}%
\end{figure}

\textit{Results}.- We start the evaluation of Eq. \ref{eco} with $h=0$ (no
disorder), increasing the interaction strength $\Delta$ from zero, see Fig.
\ref{Fig. ECOS} (\textbf{a}) and (\textbf{b}). The LE short-time scaling is
$1-M_{1,1}(t)\propto t^{4}$, which can be analytically verified by expanding
the evolution operators up to fourth order. Beyond the short-time window, it
has a smooth decay produced by the nonreversed terms $\hat{\Sigma}$
\cite{nosotros2012}. As we do not have an explicit functional form for this
regime, we define an effective characteristic time $\tau$ for the plotted
curves as the decay up to $2/3$. The rates $1/\tau$ are plotted in Fig.
\ref{Fig. Rates} as function of $\Delta$, for different disorder magnitudes $h$.

Beyond the decay regime, as shown in Fig. \ref{Fig. ECOS}-(\textbf{a}), the LE
saturates at $1/N$, which means that the initial polarization (local
excitation defined by Eq. \ref{neqstate}) is uniformly spread over the whole
spin set. This is indeed the standard picture of decoherence process leading
to an irreversible spread. But, if $\Delta$ is further increased, the LE-decay
slows down showing long tails, see Fig. \ref{Fig. ECOS}-(\textbf{b}). These
destroy the saturation at least in the time-window analyzed in the present
work. Such regime may be associated with a \textit{glassy} polarization
dynamics, i.e. the prevalence of the \textit{freezing} effect of the Ising
terms over the spreading induced by the $XY$ ones.

If we consider $\Delta=0$ and let the disorder $h$ increase, the picture is
indeed the standard Anderson localization problem. In such a case, the
localization length must be compared to the finite size of the system. Hence,
for very weak disorder, the LE degrades smoothly, as the localization length
is longer than the system's size. When the disorder is strong enough, the
localization length is smaller than the system's size and thus the initial
local excitation remains around the site $1$. In fact, the crossover between
these two physical situations can be quantified equating the system's size
with the localization length $\lambda\simeq24J^{2}/h^{2}$, given in a FGR
estimation \cite{MacKinnon93}. This yields $h=\sqrt{2}J$, in fairly
good agreement with the behavior shown in Figs. \ref{Fig. ECOS} (\textbf{c})
and (\textbf{d}).

Notice that in any case, neither Ising interactions nor Anderson disorder, can
produce a LE-decay faster than a well defined time scale (see Figs.
\ref{Fig. ECOS} (\textbf{b}) and (\textbf{d})). This is specifically
determined by the SP of the local excitation under the evolution given by
$\hat{H}_{0}$,%
\begin{equation}
P_{1,1}^{0}(2t)=2\left\langle \Psi_{neq}\right\vert \exp\left(  \mathrm{i}%
\hat{H}_{0}t/\hbar\right)  \hat{S}_{1}^{z}\exp\left(  -\mathrm{i}\hat{H}%
_{0}t/\hbar\right)  \left\vert \Psi_{neq}\right\rangle ,\label{SP}%
\end{equation}
where we emphasize that there is no dependence on $\hat{\Sigma}$. Such an
$\hat{H}_{0}$-controlled decay resembles the perturbation independent decay
experimentally observed in spin systems \cite{patricia98} and the Lyapunov
regime of classically chaotic systems \cite{jalpa}.%

In Fig. \ref{Fig. Rates} we show the role of the time-scale determined by Eq.
\ref{SP}, acting as the limit for LE decay rates. These results are quite
general, similar behavior is obtained for next nearest neighbors interactions
(both in $\hat{H}_{0}$ and $\hat{\Sigma}$), provided that $\hat{\Sigma}$ has
only Ising terms or Anderson disorder.

In order to analyze the ergodicity of the polarization dynamics observed in
our finite system, we evaluate the mean LE, $\bar{M}_{1,1}$:%
\[
\bar{M}_{1,1}(T)=\frac{1}{T}\int_{0}^{T}M_{1,1}(t)dt.
\]

\begin{figure}
    \centering
     \includegraphics[width=0.5\textwidth]{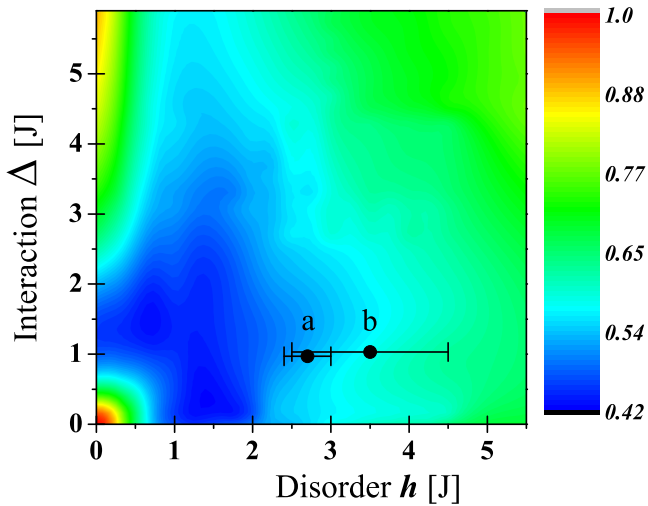}\\
      \caption{Color online. Phase diagram for $\bar{M}_{1,1}(T)$ at $T=12\hbar/J$. Data point
(\textbf{a}) is given for the MBL transition in Ref. \cite{DeLuca2012},
$\Delta=1.0J$, $h_{c}=(2.7\pm0.3)J$. Data point (\textbf{b}) is given for the MBL
transition in Ref. \cite{palhuse2010}, $\Delta=1.0J$, $h_{c}=(3.5\pm1.0)J$. These
points are slightly shifted in the plot from $\Delta=1.0J$ in order to avoid
their overlap.}%
    \label{MAPA}%
\end{figure}

The standard analysis of the Anderson localization problem should imply, in
the present case, to compute $\lim_{T\rightarrow\infty}$ $\bar{M}_{1,1}(T)$.
Instead, we analyze $\bar{M}_{1,1}(T)$ at $T=12\hbar/J$ which, as shown in
Fig. \ref{Fig. ECOS}, is long enough to allow for a uniform spreading of the
polarization, provided that $\Delta$ is strong. Also, a rigorous upper bound
for the integration time $T$ must be the system's Heisenberg time $T_{H}$, at
which finite-size recurrences show up. In Fig. \ref{MAPA} we show a level plot
of $\bar{M}_{1,1}$ as a function of the interaction $\Delta$ and disorder
strength $h$. This results in a qualitative phase-diagram which evidences
the competition between such physical magnitudes. When both $\Delta$ and $h$
are weak, $\bar{M}_{1,1}$ remains near $1$, since the system is almost
reversible. Thus, the parametric region at the bottom left corner may
be associated with \textit{decoherence}, i.e. the system is weakly perturbed
by uncontrolled degrees of freedom. If either $\Delta$ or $h$ are further
increased, the system enters in a diffusive regime where the initial local
polarization rapidly spreads irreversibly all across the spin system.
Consistently, this bluish region is associated with an \textit{ergodic}
behavior for the polarization, since it is equally distributed along the spin
system. For low disorder ($h\lesssim0.5J$), increasing $\Delta$ leads to an
Ising predominance, which \textit{freezes} the polarization dynamics. We
interpret such behavior as a \textit{glassy} dynamics, with long relaxation
times. This localization keeps $\bar{M}_{1,1}$ high. Analogously, increasing
$h$ for a fixed value of $\Delta$ evidences a \textit{smooth crossover} to a
localized phase, where the polarization does not diffuse considerably.

Notice that localization by disorder is weakened when $1.0J\lesssim
\Delta\lesssim2.0J$, since the ergodic region seems to unfold for larger $h$.
In fact, such values of interaction strength correspond to a faster arrival to
the $1/N$ saturation, as shown in Figs. \ref{Fig. ECOS} (\textbf{a}) and
(\textbf{b}) (when $h=0$). Additionally, data around the $\Delta$ axis show
that the disorder tends to \textit{abruptly} destroy the quenching produced by
Ising interactions. This seems suggestive of a parameter region where the
interaction-disorder competition leads to a sharp transition between glassy
and ergodic phases. However, a reliable finite size scaling of this regime
would require excessively long times to capture how a vitreous dynamics is
affected by disorder.

Previous numerical results of the SP \cite{DeLuca2012} and an analysis of the
many-body eigenstates \cite{palhuse2010} of the same spin model have
identified critical values for the MBL transition. Quite remarkably, they lie
precisely at the crossover between the ergodic and the localized phases of the
LE; see data-points (\textbf{a}) and (\textbf{b}) in Fig. \ref{MAPA}. In our
simulations, increasing $N$ (e.g. 10, 12 and 14) enables a larger integration
time $T$, since $T\lesssim T_{H}\propto N$. In fact, when $\Delta\sim1.0J$, it
can be verified that both sides of the transition are well behaved since
$\bar{M}_{1,1}\sim1/N$ in the ergodic regime, while $\bar{M}_{1,1}%
\sim1/\lambda$ for $h$ strong enough (regardless of $N$), and $d\bar{M}%
_{1,1}/dh$ increases with $N$. However, this finite size scaling of $\bar
{M}_{1,1}(T)$ within our accessible range is not enough to yield precise
critical values for the MBL transition.

In summary, we were able to draw a qualitative phase-diagram for the
polarization under the LE dynamics, identifying ergodic, localized and glassy
regimes. It displays a nontrivial geography with a deep penetration of the
ergodic phase into the glassy domain separating it from the localized region.
Besides, while in finite 1-d systems Anderson localization is indeed a smooth
crossover, it seems to develop into a ergodic-localized transition for nonzero
interactions. Additionally, our results suggest that the glassy-ergodic
transition is a better candidate for a sharp phase transition. In spite of the
fact that the local nature of the observable constitutes a limitation to
perform a reliable finite size scaling, our strategy seems promising to
analyze different underlying topologies and different ways to breakdown
integrability. Last, but not least, in state-of-the-art nuclear
magnetic resonance \cite{Franzoni2012,Boutis2012}, the high temperature correlation functions, like the LE, 
are recognized witnesses for the onset of phase transitions\cite{Laflamme2009LE}, even hinting at the appearance of 
many-body localization \cite{Franzoni2005,AlvarezLocalizacion}.%

P.R.Z., A.D.D. and H.M.P. wish to dedicate this paper to the memory of their coauthor
P.R.L. who did not live to see the final version of this paper. This work benefited from discussions with L.F.
Santos, and comments by O. Osenda and F. Pastawski. P.R.Z. and H.M.P. acknowledge M.C. Ba\~{n}uls and J.I. Cirac for their kind hospitality at MPQ in Garching. We acknowledge support from CONICET, ANPCyT, SeCyT-UNC and MinCyT-Cor. The calculations were
done on Graphical Processing Units under a NVIDIA Professor Partnership Program led by O. Reula.

%

\end{document}